\documentclass[floats,floatfix,amssymb,prd,twocolumn,superscriptaddress,nofootinbib]{revtex4-2}

\usepackage[english]{babel}
\usepackage[utf8]{inputenc}
\usepackage{amsmath}
\usepackage{mathbbol}
\usepackage{amssymb}
\usepackage{bbold}
\usepackage{graphicx,amsfonts}
\usepackage{epsfig}
\usepackage[svgnames]{xcolor}
\definecolor{crimsonglory}{rgb}{0.75,0.0,0.2}
\usepackage[colorlinks=true,
linkcolor=crimsonglory,
urlcolor=crimsonglory,
citecolor=crimsonglory]{hyperref}
\usepackage{bm}
\usepackage{mathrsfs}
\usepackage{mathtools}
\usepackage{enumerate}
\usepackage{amsthm}
\usepackage{bbm}
\usepackage{comment}
\usepackage{physics}
\usepackage{url}
\usepackage{upgreek}
\usepackage[title]{appendix}
\usepackage{float} 
\usepackage{rotating}
\usepackage{booktabs}

\providecommand{\abs}[1]{\lvert#1\rvert}

\newcommand{\be}{\begin{eqnarray}}
\newcommand{\ee}{\end{eqnarray}}
\newcommand{\bea}{\begin{eqnarray}}

\newcommand{\eea}{\end{eqnarray}}

\makeindex

\begin{document}

\title{Near-horizon Geodesic Instabilities and Anomalous Decay of Quasinormal Modes in Weyl Black Holes
}

\author{Gerasimos Kouniatalis}
\email{gkouniatalis@noa.gr}
\affiliation{Physics Division, School of Applied Mathematical and Physical Sciences, National Technical University of Athens, 15780 Zografou Campus,
    Athens, Greece.}
    \affiliation{National Observatory of Athens, Lofos Nymfon, 11852 Athens, 
Greece}

\author{P. A. Gonz\'{a}lez}
\email{pablo.gonzalez@udp.cl} \affiliation{Facultad de
Ingenier\'{i}a y Ciencias, Universidad Diego Portales, Avenida Ej\'{e}rcito
Libertador 441, Casilla 298-V, Santiago, Chile.}

\author{Eleftherios Papantonopoulos}
\email{lpapa@central.ntua.gr}
\affiliation{Physics Division, School of Applied Mathematical and Physical Sciences, National Technical University of Athens, 15780 Zografou Campus,
    Athens, Greece.}

\author{Yerko V\'asquez}
\email{yvasquez@userena.cl}
\affiliation{Departamento de F\'isica, Facultad de Ciencias, Universidad de La Serena,\\
Avenida Cisternas 1200, La Serena, Chile.}


\begin{abstract}

We study the stability of the Weyl geometry considering an exact black hole solution. By calculating the geodesics of massless and massive scalar fields orbiting outside the Weyl black hole background and using the Lyapunov exponent, we show that geodesic instabilities,  characterized by the Lyapunov exponent,
appear in the asymptotically de Siter-like spacetime. Calculating the photon sphere's quasinormal modes (QNMs) of a scalar field perturbing the Weyl black hole, we find a relation connecting the QNMs with the Lyapunov exponent in the asymptotically de Siter-like spacetime. Furthermore, we study the anomalous decay rate of the QNMs connecting their behavior with the Lyapunov exponent. 

\end{abstract}

\maketitle

\section{Introduction}

Weyl geometry \cite{Weylgeom1,Weylgeom2} is a generalization of Riemannian geometry in which gravity and electromagnetism are unified. This gravitational theory is conformal invariant having a nonmetric geometry with the covariant derivative of the metric tensor being proportional to a vector field. Dirac \cite{Dirac1,Dirac2} proposed a generalization of  Weyl's theory by introducing a real scalar field.  Cosmological applications of Weyl theory in the presence of a scalar field were considered in detail in \cite{Dirac3}, \cite{Dirac4}, and \cite{cosm} and further generalizations of Weyl theory were considered in \cite{Utiyama1,Utiyama2,Nishioka, Gerasimos}.

In Weyl geometric gravity theory black hole type solutions in spherical symmetry were investigated. One of the first exact vacuum solutions of Weyl gravity theory, given by
 $
 A(r)=1-3\beta \gamma -\frac{\beta\left(2-3\beta \gamma\right)}{r}+\gamma r+kr^2,
 $
 where $\beta$, $\gamma$ and $k$ are constants, was found in \cite{M1}. A metric similar in form to the exact Weyl gravity vacuum solution was found in \cite{Pi} as a solution of the field equations of dRGT massive gravity theory. Additionally, black hole type solutions in spherical symmetry were investigated in detail in \cite{HT3}, by using numerical and analytical methods. 

The study of motion of massive and massless particles following geodesics around black holes may give us important information on the background geometry reveling its structure. Circular geodesics are particularly interesting, allowing us to study astrophysical events such as gravitational binding energy and astrophysical black
holes. In \cite{Cardoso:2008bp} a detailed study of null geodesics was carried out. Unstable circular null geodesics are generated by the gravitational collapse of stars \cite{podurets,amesthorne}.

Another very important tool in understanding the properties of compact objects and distinguishing their nature is the knowledge of the quasinormal modes (QNMs) and quasinormal frequencies (QNFs). The QNMs give an infinite discrete spectrum consisting of complex frequencies, $\omega = \omega_R + i\omega_I$, where the real part $\omega_R$ determines the oscillation timescale of the modes, while the complex part $\omega_I$ determines their exponential decaying timescale (for a review on QNMs see \cite{Kokkotas:1999bd,Berti:2009kk, Konoplya:2011qq} and recent works \cite{Panotopoulos:2020mii,Rincon:2020pne}). The QNMs and QNFs can give us important information about the stability of matter fields that evolve perturbatively in the exterior region of a compact object without backreacting on the metric. 
The complex QNM frequencies are determined by the angular velocity at the unstable
null geodesic, whereas the imaginary part is related to the instability timescale of
the orbit. It was found that for the Schwarzschild and Kerr black hole background the longest-lived gravitational modes are always the ones with a lower angular number. This is expected because the more energetic modes with a high angular number $\ell$ would have faster decaying rates.

Recent works in Weyl gravity show a richer structure in quasinormal ringing beyond the standard Schwarzschild behavior. For example, \cite{Konoplya:2020fwg} studies ``two stages'' of ringing: after the standard Schwarzschild-like stage, there is a long-lived dark matter inspired stage before the exponential tail in conformal Weyl gravity metrics. Additionally, newer results show that in dark matter-inspired Weyl gravity black holes and wormholes, the quasinormal spectra fall into several branches, and shadows of these objects differ, allowing for potential observational distinction \cite{Konoplya:2025mvj}. Long-lived QNMs and gray-body factors for massive scalar field perturbations have been computed in Weyl gravity for black holes and wormholes, showing that the presence of mass or ``dark matter''-type terms can extend the lifetime of certain modes significantly \cite{Lutfuoglu:2025hjy}.

In \cite{Cardoso:2008bp} there is a detailed study of the relation between unstable null geodesics, Lyapunov exponents, and quasinormal modes. A formula was found that connects the Lyapunov
exponent $\lambda$ with the QNMs of unstable circular null geodesics for any static, spherically symmetric, asymptotically flat
spacetime. In this result it was found that it is valid for a wide class of spacetimes and geodesics, including stationary spherically symmetric spacetimes and equatorial orbits in the geometry of higher-dimensional rotating Myers-Perry black hole solutions \cite{Myers}. However, it was shown that the link between null geodesics and QNMs is violated in Einstein-Lovelock theory \cite{Konoplya:2017wot}. Furthermore, the relation between null geodesics and QNMs has been further clarified in \cite{Konoplya:2022gjp}: In some scenarios, the usual correspondence holds in the eikonal limit, but deviations can occur beyond it. 
A comprehensive approach for the derivation of analytical expressions for QNMs and gray-body factors at various orders beyond the eikonal limit has been provided in \cite{Konoplya:2023moy}

There are many studies conducted on black hole spacetime perturbations and around particle orbits \cite{Bombelli:1991eg}, which are nonlinear and nonintegrable in the general chaos theory. The Lyapunov exponent can be used in understanding the separation rate between neighboring trajectories, which reflects the sensibility of the system to the initial condition. The positive Lyapunov exponent indicates that if there is initially a slight divergence of the geodesics will lead to exponential separation of trajectories.  When the Lyapunov exponent is $\lambda =0$, the neighboring trajectories will neither diverge or converge. For $\lambda <0$, the particle orbit will be asymptotic stable, meaning that the nearby trajectories will tend to overlap. Outside the horizon of a black hole the Lyapunov exponent can be used to probe the orbits stability and rate of orbits divergence of the massive and massless particles. The information gained by the Lyapunov exponent has already been investigated in Schwarzschild-Melvin spacetime \cite{Wang:2016wcj}, accelerating and rotating black holes \cite{Chen:2016tmr} and Born-Infeld AdS spacetime \cite{Yang:2023hci}.

If we perturb a background black hole with a massless scalar field, the longest-lived modes are the ones with a higher angular number $\ell$. However,  if the perturbed scalar field is massive, then there is a critical mass of the scalar field where the behavior of the decay rate of the QNMs is inverted and then the longest-lived modes are the ones with lower angular number. This can be understood from the fact that massive scalar fields offer more energy in the perturbed system. This anomalous behavior in the QNFs is possible in asymptotically flat, in asymptotically dS and in asymptotically AdS spacetimes. However, it was shown that the critical mass exists for asymptotically flat and asymptotically dS spacetimes, and it is not present in asymptotically AdS spacetimes for large and intermediate black holes.
This behavior has been extensively studied for scalar fields \cite{Konoplya:2006br,Dolan:2007mj, Lagos:2020oek,      Aragon:2020tvq,Aragon:2020xtm,Fontana:2020syy,Becar:2023jtd, Becar:2023zbl, Becar:2024agj, Becar:2025niq} as well as charged scalar fields \cite{Gonzalez:2022upu,Becar:2022wcj} and fermionic fields \cite{Aragon:2020teq}, in black hole spacetimes. The anomalous decay in accelerating black holes was studied in \cite{Destounis:2020pjk}. Furthermore, it has been recently studied for scalar fields in wormhole spacetimes \cite{Gonzalez:2022ote,Alfaro:2024tdr}.

In this work we will study the stability of the Weyl geometry considering a specific black hole solution discussed in \cite{Yang:2022icz,Sakti:2024pze}. We will study the geodesics of massless and massive test particles orbiting outside the background Weyl black hole. Using the Lyapunov exponent we will show that the chaotic instabilies appear in the asymptotically de Siter-like spacetime, and we will constrain the values of the vector field and the scalar field appearing in the background black hole. Calculating the QNMs of a test scalar field perturbing the Weyl black hole, we will find a relation connecting the QNMs with the Lyapunov exponent of the unstable circular orbit. We will also study the connection of the Lyapunov exponent with the anomalous decay of QNMs.

In the context of conformal Weyl gravity, Ref.~\cite{Momennia:2019cfd} analyzed scalar perturbations of near-extremal Weyl black holes and obtained an analytical expression for the quasinormal frequencies. Their results showed that the real and imaginary parts of the frequencies are determined by the angular velocity and the Lyapunov exponent of the unstable circular orbit, confirming the validity of the eikonal correspondence in conformal gravity.

The work is organized as follows. In Section \ref{app:1} we give a general description of the Weyl conformal geometry. In Section \ref{app:11} we discuss the Weyl black hole solution we used. In Section \ref{geodesics} we study massless and massive particle geodesics. In Section \ref{black} we study the photon sphere's QNMs for asymptotically dS-like black holes. In Section 
\ref{CLmc} we connect the Lyapunov exponent with the anomalous behavior.
Finally, 
in Section \ref{conc} are our conclusions.

\section{The Weyl Geometry}\label{app:1}

In this section, we give a general description of the Weyl conformal geometry. A detailed presentation of the Weyl geometry is given in \cite{Yang:2022icz}.
The Weyl conformal geometry is defined as the equivalence classes of ($g_{\mu\nu}, \omega_\mu$) of the metric
and of the Weyl gauge field ($\omega_\mu$), related by the Weyl gauge transformations
\begin{eqnarray}
\notag \hat{g}_{\mu\nu}&=&\!\Sigma^d 
\,g_{\mu\nu}\,,\\
\notag \sqrt{-\hat{g}}&=&\Sigma^{2 d} \sqrt{-g}\,,\\ 
\hat\omega_\mu&=&\omega _\mu -\frac{1}{\alpha}\, \partial_\mu\ln\Sigma \,,
\label{W1}
\end{eqnarray}
where $d$ is the Weyl weight (charge) of  $g_{\mu\nu}$,  
$\!\Sigma$  is the conformal factor,  while $\alpha$ is the Weyl gauge coupling. For simplicity reasons, it is  considered $d=1$. The Weyl gauge field is connected with the Weyl connection $\tilde\Gamma$,
\begin{equation}
\label{W2}
\tilde\nabla_\lambda  g_{\mu\nu}=\partial_\lambda g_{\mu\nu}- \tilde \Gamma^\rho_{\mu\lambda}
g_{\rho\nu} -\tilde\Gamma^\rho_{\nu\lambda}\,g_{\rho\mu}=- d\, \alpha\, \omega _{\lambda} g_{\mu\nu}~,
\end{equation}
which indicates that the Weyl geometry is {\it non-metric}.  Therefore,  in Weyl geometry  the covariant derivatives $\nabla_\lambda$,  acting on the geometric and physical quantities, are replaced by their Weyl-geometric counterpart. The expression of $\tilde \Gamma$ in  Eq.~(\ref{W2}) is given by 
\bea\label{AGamma}
\tilde \Gamma_{\mu\nu}^\lambda=
\Gamma_{\mu\nu}^\lambda+\alpha \,\frac{d}{2} \,\Big[\delta_\mu^\lambda\,\omega _\nu
+\delta_\nu^\lambda\, \omega _\mu - g_{\mu\nu} \,\omega ^\lambda\Big]~,
\eea
and taking its trace 
 and denoting $\Gamma_\mu\equiv \Gamma_{\mu\lambda}^\lambda$ and
 $\tilde\Gamma_\mu\equiv \tilde\Gamma_{\mu\lambda}^\lambda$, respectively, we obtain
\bea
\tilde \Gamma_\mu=\Gamma_\mu + 2 d \,\alpha\,\omega _\mu~.
\eea
As we can see from the above relation, the Weyl gauge field can be interpreted as describing the deviation of  the Weyl connection from the Levi-Civita connection $\Gamma_{\mu\nu}^\lambda$. Also it is important to  note that  $\tilde \Gamma$  is invariant under the group of conformal transformations (\ref{W1}). 

An important property of Weyl geometry is that $\tilde R$ transforms covariantly under the transformations (\ref{W1})
\bea
\hat{\tilde R}=(1/\Sigma^d)\,\tilde R~,
\eea
and then, it follows immediately that the term $\sqrt{g}\, \tilde R^2$
is also Weyl gauge invariant.

A geometrical quantity that is important in Weyl geometry is the strength of the Weyl vector field $\tilde{F}_{\mu \nu}$, defined as
\be
\tilde{F}_{\mu \nu}=\nabla _\mu \omega _\nu-\nabla _\nu \omega _\mu~.
\ee
Considering a conformally invariant gravitational Lagrangian density
\begin{equation}
L_{W}=\left( \frac{1}{4!\,\xi ^{2}}\tilde{R}^{2}-\frac{1}{4}\tilde{F}_{\mu
\nu }\tilde{F}^{\mu \nu }\right) ,  \label{inA}
\end{equation}%
where  $\xi <1$ is the parameter of the perturbative
coupling,  the action of the Weyl geometric gravitational theory can be obtained by linearizing the action built from the Lagrangian $L_W$ in Eq. (\ref{inA}) (\cite{Yang:2022icz, Sakti:2024pze})
\begin{eqnarray}
\mathcal{S} &=&\int \Bigg[-\frac{\phi ^{2}}{12\xi ^{2}}\Big(R-3\alpha
\nabla _{\mu }\omega ^{\mu }-\frac{3}{2}\alpha ^{2}\omega _{\mu }\omega
^{\mu }\Big)  \notag \\
&&-\frac{\phi ^{4}}{4!\,\xi ^{2}}-\frac{1}{4}\tilde{F}_{\mu \nu }\tilde{F%
}^{\mu \nu }\Bigg]\sqrt{-g}\,d^{4}x~. \label{actiong}
\end{eqnarray}
where $\phi$ is an auxiliary scalar.

\section{Weyl black-hole metric}\label{app:11}

Using the action (\ref{actiong}) we will review the  black hole solution arising from this action as it was derived in \cite{Yang:2022icz, Sakti:2024pze}.
It was considered a static spherically symmetric configuration, with the metric given in a general form by
\begin{equation}\label{eq:spherical}\
	ds^2 = e^{\nu(r)} dt^2 -e^{\mu(r)}dr^2 - r^2 d\Omega^2~,
\end{equation}
where $d\Omega^2 = d\theta^2 +\sin^2\theta d\varphi^2$. The Weyl vector field $\omega _\mu$, was represented as $\omega_\mu =(\omega_0, \omega_1,0, 0)$ and they assumed that $\omega_0=0$.
From the assumption of the form on the Weyl vector it follows that $F_{\mu\nu}\equiv 0$. Then,  by employing this condition from the Weyl vector field equation of motion, they obtained the following result
\begin{equation}
\Phi'=\alpha\Phi\omega_1~,\label{eq:simplifiedKGEq}\
\end{equation}
where $\Phi =\phi^2$, and the prime symbol (${}^\prime$) denotes differentiation with respect to $r$.

Considering the field equations 
\begin{equation}
\Box \Phi = \frac{1}{\sqrt{-g}}\frac{\partial}{\partial x^\mu}\left(\sqrt{-g}g^{\mu\nu}\frac{\partial\Phi}{\partial x^\nu} \right)~,
\end{equation}
and
\begin{equation}
	\nabla_\mu \omega^\mu = \frac{1}{\sqrt{-g}}\frac{\partial}{\partial x^\mu}\left(\sqrt{-g}\omega^\mu \right)~,
\end{equation}
and defining the effective energy density $\rho$ and pressure $p$ associated to the scalar field and to the Weyl geometric function, one can find black hole solutions in Weyl geometric gravity by assuming $g_{tt}g_{rr}\neq-1$. Then, by writing 
\begin{equation}
\nu (r) +\mu (r) = h(r)~,\label{eq:nulambdaf}
\end{equation}
where $h(r)$ is an arbitrary function of the radial coordinate, as a function of the scalar field $\Phi$, the function $h(r)$ can be found as 
\begin{equation}\label{eq:f(r)integral}
h(r) = \int \frac{2\Phi'' \Phi - 3\Phi'^2}{2\Phi+r\Phi'} rdr~.
\end{equation}
Then, a black hole solution was found in \cite{Yang:2022icz} which corresponds to the case when $g_{tt}g_{rr}=-1$. The following condition for the metric tensor potentials was considered
\begin{equation}
\nu(r) +\mu(r) = 0,~~ \forall r>0~.\label{eq:lambdaplusnu}\
\end{equation}
Then, the differential equation satisfied by $\Phi$ was found
\begin{equation}
\Phi''=\frac{3\Phi'^2}{2\Phi~},
\end{equation}
corresponding to the choice $h(r)=0$ in Eq.~(\ref{eq:f(r)integral}). The solution of the above equation yields
\begin{equation}
\Phi(r)=\frac{C_1}{(r+C_2)^2}~,\label{eq:scalarsol1}\
\end{equation}
where $C_2$ is just an arbitrary integration constant. The scalar field satisfies the condition $\Phi(r) \rightarrow 0$ at infinity. One can also find the Weyl vector field
\begin{equation}
\omega_1 = \frac{\Phi'}{\alpha \Phi}=-\frac{2}{\alpha(r+C_2)}~.\label{eq:Weylvectorsol1}\
\end{equation}
Then, from the gravitational field equation one can find the metric potentials as
\begin{equation}
e^{-\mu}=e^{\nu}=1-\delta+\frac{\delta(2-\delta)}{3r_g}r -\frac{r_g}{r}+C_3r^2~, \label{eq:StarType1}
\end{equation}
where $\delta, r_g, $ and $C_3$ are arbitrary constants and $C_2 = 3r_g/\delta$.

This metric is the generalization of the Schwarzschild-de Sitter
solution. If $C_3 = 0$, the resulting metric will mimic spacetime in GR but
with additional linear term in $r$. For $\delta = 0$ the spacetime will become
asymptotically flat. However, the case $\delta =2$ is excluded from our analysis
because it leads to singular behavior in the circular orbit equations.
Specifically, the photon sphere radius $r_{ps}$ diverges as $\delta \rightarrow 2$, making
the Lyapunov exponent analysis invalid. Furthermore, the metric develops
signature issues as discussed in Section IV. For negative $C_3$, there will
be cosmological horizon and the spacetime will be asymptotically de
Sitter-like. For positive $C_3$, the spacetime could become asymptotically
Anti-de Sitter. 


\section{Massless and massive particle geodesics}
\label{geodesics}

We aim to study the stability of this theory. To this end, we calculate the Lyapunov exponent based on this setup. 
%
%
We begin by employing the Lagrangian formalism for geodesic motion \cite{Chandrasekhar:579245}. In the spherically symmetric spacetime described by Eq. (\ref{eq:spherical}), the Lagrangian may be expressed as
\begin{equation}
2\mathcal{L} = e^{\nu(r)}\Dot{t}^2 -  e^{\mu(r)}\Dot{r}^2 -r^2(\Dot{\theta}^2 + \sin^2{\theta} \Dot{\varphi}^2)~,
\end{equation}
where the over-dot denotes differentiation with respect to the affine parameter $\sigma$ of the geodesic.

From this Lagrangian one obtains the canonical momenta 
\begin{equation}
    \begin{aligned}
        p_t =& \frac{\partial \mathcal{L}}{\partial \dot{t}} = e^{\nu(r)}\dot{t}~, \\
         p_r =& \frac{\partial \mathcal{L}}{\partial \dot{r}} =-e^{\mu(r)}\dot{r}~, \\
          p_{\theta} =& \frac{\partial \mathcal{L}}{\partial \dot{\theta}} = -r^2\dot{\theta}~, \\
           p_{\varphi} =& \frac{\partial \mathcal{L}}{\partial \dot{\varphi}} =- r^2\sin^2{\theta}\dot{\varphi}~. \\
    \end{aligned}
\end{equation}
 The time component of the  geodesic equations yields the conserved energy, and the $\varphi$ component yields the conserved azimuthal angular momentum (for massive particles these correspond to energy and angular momentum per unit mass, respectively) 

\begin{equation}
    \frac{d}{d\sigma} \left( \frac{\partial \mathcal{L}}{\partial \Dot{t}}~ \right) =  \frac{\partial \mathcal{L}}{\partial t} =0 \Rightarrow E\equiv p_t = e^{\nu(r)}\Dot{t}~.
\end{equation}

\begin{equation}
    \frac{d}{d\sigma} \left( \frac{\partial \mathcal{L}}{\partial \Dot{\varphi}} \right) =  \frac{\partial \mathcal{L}}{\partial \varphi} =0 \Rightarrow  L_z \equiv - p_{\varphi} = r^2\sin^2{\theta}\Dot{\varphi}
\end{equation}

The radial component of the equations of motion produces 
\begin{equation}
    r^2 \ddot{\theta} + 2r\dot{r}\dot{\theta} = r^2\sin{\theta}\cos{\theta}\dot{\varphi}^2~,
\end{equation}
therefore, choosing $\theta = \pi/2$ when $\dot{\theta}=0$ it gives us $\ddot{\theta}=0$. 
Thus, the orbit is confined to the equatorial plane $\theta = \pi/2$
\begin{equation}
    2\mathcal{L} = \frac{E^2}{e^{\nu(r)}} - \frac{\dot{r}^2}{e^{-\mu(r)}} - \frac{L^2}{r^2} = m^2~, 
\end{equation}
and therefore 
\begin{equation} \label{Potential}
    \dot{r}^2 \equiv V_r = \left(\frac{E^2}{e^{\nu(r)}}  - \frac{L^2}{r^2}  -m^2 \right)e^{-\mu(r)}~.
\end{equation}

To compute the Lyapunov exponents $\lambda$, we work in the phase space $(r, p_r)$ by linearizing the equations of motion about the circular orbit: 
\begin{equation}
    \begin{aligned}
        p_r =& e^{\mu(r)}\dot{r} \Rightarrow \\
        \dot{r} =& e^{-\mu(r)}p_r \Rightarrow \\
        \frac{d}{dr} \delta r =& \frac{e^{-\mu(r)}}{\dot{t}}\delta p_r,
    \end{aligned}
\end{equation}
from the equations of motion, we have
\begin{equation}
    \begin{aligned}
        \frac{d p_r}{d\sigma} =& \frac{\partial \mathcal{L}}{\partial r} \Rightarrow \\
        \frac{d}{dt}\delta p_r =& \frac{1}{\dot{t}} \frac{d}{dr} \left(\frac{\partial \mathcal{L}}{\partial r}  \right)\delta r~.
    \end{aligned}
\end{equation}
For circular motion of particles (where $p_r=0$), the Jacobian matrix reduces to
 \begin{equation}
     K_{ij}= \begin{pmatrix}
0 & K_1\\
K_2 & 0 
\end{pmatrix}\,,
 \end{equation}
 where 
\begin{equation}
    K_1 = \frac{1}{\dot{t}} \frac{d}{dr} \left(\frac{\partial \mathcal{L}}{\partial r}  \right)~,
\end{equation}
and 
\begin{equation}
    K_2 = \frac{1}{e^{\mu(r)}\dot{t}}~,
\end{equation}
which implies 
\begin{equation}
    \lambda = \pm \sqrt{K_1K_2}~.
\end{equation}
Since for circular orbits one has  $\dot{r}=0\Rightarrow V_r=V'_r=0$,
the Lyapunov exponent may be defined in terms of the second derivative of the effective potential governing radial motion $V_r$ \cite{Cardoso:2008bp}
\begin{equation}
\label{sle}
    \begin{aligned}
        \lambda = \pm \sqrt{\frac{V_r''}{2\dot{t}^2}}~.
    \end{aligned}
\end{equation}

The Lyapunov exponent is valid for many spacetimes and geodesics, including stationary spherically symmetric spacetimes and equatorial orbits in the geometry of higher-dimensional rotating  black hole solutions.

\subsection{Massless particles}

For null geodesics (i.e. $m^2=0 $) we have 
\begin{equation}
    \begin{aligned}
        V_r = 0 \Rightarrow \frac{E}{L} = \pm \sqrt{\frac{e^{\nu_c}}{r_c^2}}~, 
    \end{aligned}
\end{equation}
where $r_c$ is the (constant) radius of the circular orbit we examine. 

In addition,
\begin{equation}
    \begin{aligned}
        V_r' = 0 \Rightarrow \frac{E}{L} = \pm \sqrt{\frac{2e^{\nu_c}}{r_c^3\nu_c'}}~,
    \end{aligned}
\end{equation}
which leads to 
\begin{equation}
    \nu'(r_c) = \frac{2}{r_c}~.
\end{equation}
We then calculate the fraction which appears in the Lyapunov exponent
\begin{equation}
\label{LE1}
    \frac{V_r''}{2\dot{t}^2} = -e^{\nu_c-\mu_c} \frac{\nu_c''r_c^2+2}{2r_c^2}\,.
\end{equation}
Therefore, instabilities are present if 
\begin{equation} \label{inequality1}
    \nu_c''r_c^2 < -2~,
\end{equation}
which, using (\ref{eq:StarType1}) reduces to a range of values at which the above relation is valid,
\begin{equation}
    \begin{aligned}
        r_c \in& (0.69731, 1.10143), 
        (1.10143, 2.82459)~, \\
        &(4.89738, 9.61827),  (9.61827, +\infty)~,
    \end{aligned}
\end{equation}
where we keep in mind that the horizons are located at $1.12$ and $10.27$ and the relation (\ref{eq:StarType1}) is valid only in this range of values for $r_c$, therefore the last range of values $(9.61827, +\infty)$ is not valid.

Furthermore, using the relation 

\begin{equation} \label{omega1}
    r=-\frac{\omega_1C_2+1}{\omega_1}~,
\end{equation}
which appears in \cite{Sakti:2024pze}, we can turn the inequality (\ref{inequality1}) into a constrain for the Weyl gauge field $\omega_1$.
Furthermore, using the relation (\ref{eq:scalarsol1})
we can also constrain the scalar field. 
Using the above relations and the range of constrains we find that 
\begin{equation}
    \omega_1 \in (-0.2,0)~,
\end{equation}
and
\begin{equation}
    \Phi \in (0,2.74)~.
\end{equation}
In these estimations we used $C_1=65.2$, $C_2=15$, $C_3=-0.02$ and $\delta=0.2$ as described in \cite{Sakti:2024pze}. We present the evolution of the radial component of the Weyl gauge field and the scalar $\Phi$ with respect to $r$ (as we move away from the black hole) in the next two figures. 
\begin{figure}[h!]  
\centering
\includegraphics[width=0.40\textwidth]{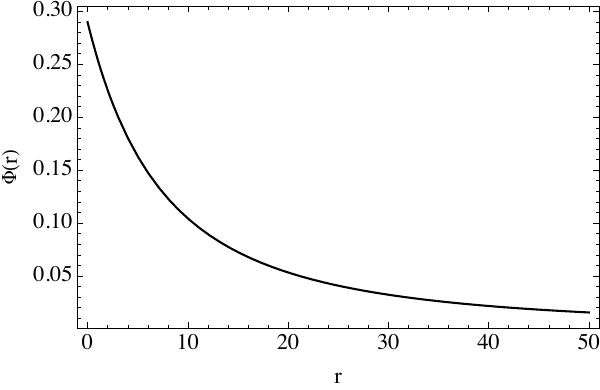}  
 \caption{Behavior of the scalar field $\Phi(r)$ as a function of the radial coordinate $r$ (in units of $r_g$, where $r_g$ is the gravitational radius). The scalar field follows the relation $\Phi(r) = C_1/(r + C_2)^2$ with $C_1 = 65.2$ and $C_2 = 15r_g$, satisfying the boundary condition $\Phi(r) \to 0$ at infinity. The plot shows the monotonic decrease of the scalar field strength as we move away from the black hole horizon.}
  \label{fig:plot1}  
\end{figure}

\begin{figure}[h!]  
  \centering
  \includegraphics[width=0.4\textwidth]{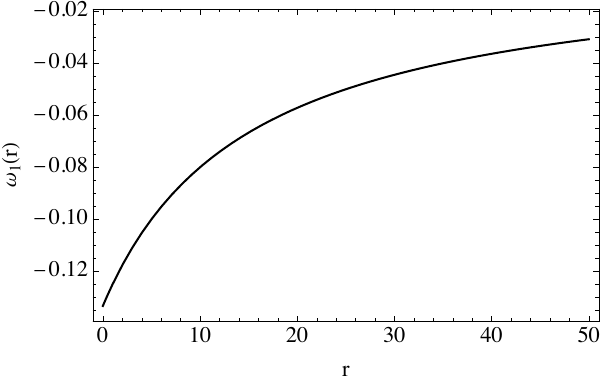}  
 \caption{Behavior of the radial component of the Weyl vector field $\omega_1(r)$ as a function of the radial coordinate $r$ (in units of $r_g$). The Weyl vector field follows the relation $\omega_1(r) = -2/[\alpha(r + C_2)]$ with $C_2 = 15r_g$, derived from the scalar field equation $\Phi' = \alpha\Phi\omega_1$. The plot shows the inverse relationship between the Weyl vector field strength and radial distance, with $\omega_1(r)$ approaching zero asymptotically far from the black hole. The field is plotted in natural units where the Weyl gauge coupling $\alpha$ sets the scale.}
  \label{fig:plot1}  
\end{figure}

\newpage

\begin{figure}[h]
	\begin{center}
		\includegraphics[width=80mm]{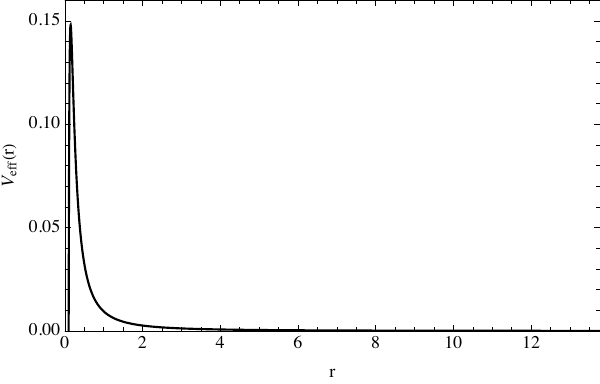}
	\end{center}
	\caption{Plot of the effective potential of photons. Here we have used the values $L=0.1$, $\delta=0.01$, $r_g=0.1$, and $C_3=-0.01$. The plot shows that the value of the photon-sphere radius 
    is $r_{ps}\approx 0.1508$, where the effective potential is maximum and it is independent of the cosmological constant. In addition, we find $r_{\Lambda}\approx 13.7701$.}
	\label{f2}
\end{figure}

In general, possible unstable circular orbits can be found studying the effective potential. Rewriting the radial equation as
\begin{equation} \label{motion}
\dot{r}^2 + V_{\text{eff}}(r) = E^2 \,.
\end{equation}
For massless particles, the effective potential $V_{\text{eff}}$ is defined by
\begin{eqnarray}\label{n1}
\notag V_{\text{eff}}\left( r\right) &=& E^2 -V_r  \\
&=& \left(1-\delta+\frac{\delta(2-\delta)}{3r_g}r -\frac{r_g}{r}+C_3r^2\right)\frac{L^2}{r^2}\,.
\end{eqnarray}
A typical graph of this effective potential is shown in Fig. \ref{f2}, where we can observe the existence of a maximum potential located at
\begin{equation}
r_{ps}= -\frac{3r_g}{\delta-2}\,,
\label{rul}
\end{equation}
which represents an unstable circular orbit, that is independent of $L$ and the constant $C_3$.

\begin{figure}[h]
	\begin{center}
		\includegraphics[width=60mm]{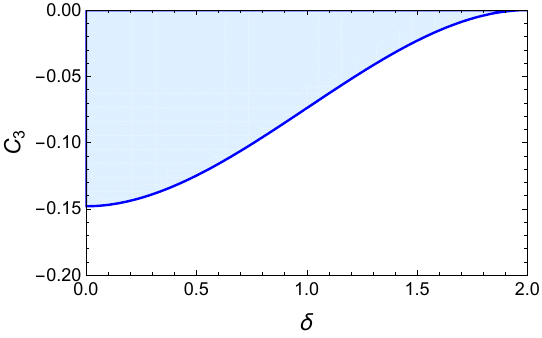}
	\end{center}
	\caption{The Lyapunov exponent $\lambda_0^2$ as a function of $C_3$, and $\delta$. Here,  $r_g=1$. }
	\label{f5}
\end{figure}

By using Eq. (\ref{sle}) and Eq. (\ref{n1}) the Lyapunov exponent for unstable circular null geodesics gives

\begin{equation}
\label{LE}
    \lambda_0^2= \frac{\xi}{27 r_g^2} \,,
\end{equation}
where $\xi=27 C_3 r_g^2+(\delta -2)^2 (\delta +1)$.
In Fig. \ref{f5} we show the region where the Lyapunov exponent is positive (shaded area), indicating a divergence between the nearby trajectories and, therefore, a strong sensitivity to initial conditions. Once a circular orbit is perturbed, the deviation grows exponentially, signaling the presence of chaos. In contrast, in the unshaded region the Lyapunov exponent is negative; however, in this case the solution does not represent a black hole.


As can be seen in Fig. \ref{f5} we have instabilities for negative values of $C_3$ for which the spacetime is asymptotically de Sitter-like. Then, using the relations (\ref{eq:scalarsol1}) and (\ref{omega1})
we can constrain the  Weyl gauge field $\omega_1$ and the scalar field $\Phi(r)$ as we have done for $C_3=-0.01$.

\subsection{Massive particles}

If we turn our interest to massive particles, we can look for timelike circular orbits along geodesics. 
To do so, we set the parameter $m^2$ in (\ref{Potential}) equal to $1$. And thus we find that 
\begin{equation}
    \begin{aligned}
        V_r = 0 \Rightarrow \frac{E^2}{e^{\nu_c}} - \frac{L^2}{r_c^2} = 1~,
    \end{aligned}
\end{equation}
and 
\begin{equation}
    \begin{aligned}
        V'_r = 0 \Rightarrow \frac{E}{L} = \pm \sqrt{\frac{2e^{\nu_c}}{r_c^3\nu'_c}}~. 
    \end{aligned}
\end{equation}
Finally, 
\begin{equation}
    \begin{aligned}
        V''_r = \left( -\frac{E^2\nu_c''}{e^{\nu_c}} +\frac{E^2\nu_c'^2}{e^{\nu_c}} - \frac{6L^2}{r_c^4} \right)e^{-\mu_c}~. 
    \end{aligned}
\end{equation}
From the first two relations, we find that 
\begin{equation}
    \begin{aligned}
        E^2 =& \frac{2e^{\nu_c}}{2-r_c\nu_c'}~, \\
        L^2 =& \frac{r_c^3\nu_c'}{2-r_c\nu_c'}~. 
    \end{aligned}
\end{equation}
We can express both the energy and the angular momentum with respect to the Weyl gauge field (or the scalar field $\Phi$). 

So, in order to find possible instabilities, the following relation must hold
\begin{equation}
    \frac{2\nu_c'^2r_c - 2\nu_c''r_c - 6\nu_c'}{r^4\nu_c'} >0~, 
\end{equation}
which is the analogue of the relation (\ref{inequality1}) we found for the photon. 
This inequality has the following solutions 
\begin{equation}
    r_c \in (0, 1.10) , (4.33, 9.62)~.
\end{equation}
Considering the second interval, since we want to be outside the horizon, we find the values of $\omega$ and $\Phi$, respectively, for which the Lyapunov exponent is positive 
\begin{equation}
    \omega_1  \in  (- 0.23, - 0.10)~,
\end{equation}
and 
\begin{equation}
    \Phi \in (0.71, 3.51)~. 
\end{equation}
The two diagrams for $\omega_1(r)$ and $\Phi(r)$ will not change, since their relations are still the same. What has changed in these diagrams are the intervals in which we see the chaotic behavior. \\

\begin{figure}[h]
	\begin{center}
		\includegraphics[width=80mm]{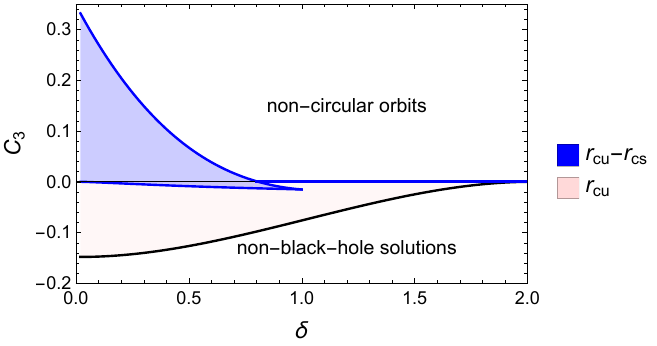}
	\end{center}
	\caption{The Lyapunov exponent for massive particles $\lambda_m^2$ as a function of $C_3$, and $\delta$. Here,  $m=1$, $r_g=1$, and $L=5$. }
	\label{LEM}
\end{figure}

For massive particles, the effective potential $V_{\text{eff}}$ is defined by
\begin{equation}\label{n2}
V_{\text{eff}}\left( r\right) = \left(1-\delta+\frac{\delta(2-\delta)}{3r_g}r -\frac{r_g}{r}+C_3r^2\right) \left( \frac{L^2}{r^2} + m^2 \right)\,,
\end{equation}
where $m^2=1$.  The location where the potential is maximum, for large values of $L$, can be estimated by
\begin{equation}
r_c\approx r_{c_0}+ \frac{r_{c_1}}{L^2}+ \frac{r_{c_2}}{L^4} + \dots \,,
\label{rul}
\end{equation}
where
\begin{eqnarray}
r_{c_0} &=&-\frac{3r_g}{\delta-2}\,, \\
r_{c_1} &=& -\frac{9 m^2 r_g^3 \xi}{(\delta -2)^5}\,, \\
r_{c_2} &=& \frac{27 m^4 \xi  \left((\delta -10) \xi +6 (\delta -2)^2\right) r_g^5}{2 (\delta -2)^9} \,,
\end{eqnarray}
which represents an unstable circular orbit. This expansion is valid for $r_g^2 \xi /(\delta-2)^4 L^2 << 1$. Now, by using Eq. (\ref{sle}) and Eq. (\ref{rul}) the Lyapunov exponent for massive particles $\lambda_m$ can be written as 

\begin{eqnarray}
\label{LEm}
\nonumber \lambda_m^2 &=& \lambda_{0}^2 \Bigg[ 1+ \frac{1}{(\delta -2)^4 L^2} \left( 81 (\delta -5) \lambda _0^2 m^2 r_g^4 \right) - \\
\nonumber && \frac{1}{2 (\delta -2)^8 L^4} \Big(729 \lambda _0^2 m^4 r_g^6 \big(2 (\delta -8) (\delta -2)^2+   \\
&& 9 ((\delta -10) \delta +40) \lambda _0^2 r_g^2\big)\Big) \Bigg] \,,
\end{eqnarray}
where $\lambda_0$ corresponds to the Lyapunov exponent for null geodesics (\ref{LE}), and coincides with $\lambda_m$ in the eikonal limit. In Fig. \ref{LEM} we plot the region where the Lyapunov exponent is positive (shadow region), which indicates a divergence between nearby trajectories, i.e., a high sensitivity to initial conditions. So, once a circular orbit is perturbed, the perturbation will increase exponentially, indicating the presence of chaos. The blue shaded region indicates the parameter values where both a stable circular orbit, with radius $r_{cs}$, and an unstable circular orbit, with radius $r_{cu}$, exist. The orange shaded region corresponds to parameter values with only an unstable circular orbit. In the unshaded region the Lyapunov exponent is negative; however, the solution does not represent a black hole solution in this region. Additionally, there is a region where there are black hole solutions, but there are no circular orbits.

\section{Photon sphere's QNMs for asymptotically dS-like black holes}
\label{black}

As we discussed in the previous section, possible instabilities appear in the asymptotically dS-like spacetime. In this section we will discuss the behavior of black holes resulting from the action (\ref{actiong}) calculating the photon sphere modes and perturbing them with a test scalar field we will study the behavior of the decay modes.

\subsection{Photon sphere modes}

The QNMs of scalar field perturbations in the background of the metric are determined by the solution to the Klein-Gordon equation
\begin{equation}
\frac{1}{\sqrt{-g}}\partial _{\mu }\left( \sqrt{-g}g^{\mu \nu }\partial_{\nu } \phi \right) =m^{2}\phi \,,  \label{KGNM}
\end{equation}%
with appropriate boundary conditions for a black hole geometry. In the above expression $m$ denotes the mass of the test scalar field $\phi$. Henceforth, we will denote the metric function (\ref{eq:StarType1}) by $f(r)$. Now, by means of the following ansatz
\begin{equation}
\phi =e^{-i\omega t} R(r) Y(\Omega) \,,\label{wave}
\end{equation}%
the Klein-Gordon equation reduces to
\begin{eqnarray}
\notag && f(r)R''(r)+\left(  f'(r)+2\frac{f(r)}{r} \right)R'(r)\\
&&+\left(\frac{\omega^2}{f(r)}-\frac{\ell (\ell+1) }{r^2}-m^{2}\right) R(r)=0\,, \label{radial}
\end{eqnarray}%
where $\ell=0,1,2,...$ represents the azimuthal quantum number and the prime denotes the derivative with respect to $r$.
Now, defining $R(r)=\frac{F(r)}{r}$
and by using the tortoise coordinate $r^*$ defined by
$dr^*=\frac{dr}{f(r)}$, the Klein-Gordon equation can be rewritten as a one-dimensional Schr\"{o}dinger equation
\begin{equation}\label{ggg}
\frac{d^{2}F(r^*)}{dr^{*2}}-V(r)F(r^*)=-\omega^{2}F(r^*)\,,
\end{equation}
where the effective potential $V(r)$, parametrically thought as $V(r^*)$, is given by
\begin{equation}\label{pot}
V(r)=f(r) \left(\frac{f'(r)}{r}+\frac{\ell (\ell+1)}{r^2} + m^2 \right)~.
\end{equation}

The QNMs via the WKB approximation are determined by the behavior of the effective potential near its maximum value $V(r^*_{max})$. The Taylor series expansion of the potential around its maximum is given by the following expression
\begin{equation} \label{expansion}
V(r^*)= V(r^*_{max})+ \sum_{k=2}^{k=\infty} \frac{V^{(k)}}{k!} (r^*-r^*_{max})^{k} \,,
\end{equation}
where
\begin{equation}
V^{(k)}= \frac{d^{k}}{d r^{*k}}V(r^*)|_{r^*=r^*_{max}}\,,
\label{eq:derivadas}
\end{equation}
corresponds to the $k$-th derivative of the potential with respect to $r^*$, evaluated at the position of the maximum of the potential, $r^*_{max}$. Using the WKB approximation up to third order beyond the eikonal limit, the QNFs are given by the following expression \cite{Iyer:1986np, Hatsuda:2019eoj}
\begin{eqnarray}
\omega^2 &=& V(r^*_{max})  -2 i U \,,
\end{eqnarray}
where
\begin{widetext}
\begin{eqnarray}
\notag U &=&  N\sqrt{-V^{(2)}/2}+\frac{i}{64} \left[ -\frac{1}{9}\frac{V^{(3)2}}{V^{(2)2}} (7+60N^2)+\frac{V^{(4)}}{V^{(2)}}(1+4 N^2) \right] +\frac{N}{2^{3/2} 288} \Bigg[ \frac{5}{24} \frac{V^{(3)4}}{(-V^{(2)})^{9/2}} (77+188N^2) + \\
\notag && \frac{3}{4} \frac{V^{(3)2} V^{(4)}}{(-V^{(2)})^{7/2}}(51+100N^2) +\frac{1}{8} \frac{V^{(4)2}}{(-
V^{(2)})^{5/2}}(67+68 N^2)+\frac{V^{(3)}V^{(5)}}{(-V^{(2)})^{5/2}}(19+28N^2)+\frac{V^{(6)}}{(-V^{(2)})^{3/2}} (5+4N^2)  \Bigg]\,,
\end{eqnarray}
\end{widetext}
and $N=n+1/2$, with $n=0,1,2,\dots$, is the overtone number.
The imaginary and real parts of the QNFs can be written as
\begin{eqnarray}
\nonumber  \label{im} Im(\omega)^2 &=& - (Im(U)+V/2)+\\
&& \sqrt{(Im(U)+V/2)^2+Re(U)^2} \,, \\
Re(\omega)^2 &=& -Re(U)^2 / Im(\omega)^2 \,,
\end{eqnarray}
respectively, where $Re(U)$ denotes the real part of $U$, and $Im(U)$ represents its imaginary part. 

Defining $L^2= \ell (\ell+1)$, we find that for large values of $L$, the maximum of the potential is approximately at
\begin{equation}
    r_{max} \approx r_{0}+\frac{r_{1}}{L^{2}}\,,
\end{equation}
where
\begin{equation}
r_{0}=-\frac{3 r_g}{\delta -2}\,,
\end{equation}
\begin{equation}
r_{1}= \frac{\xi  r_g \left[3 (\delta -2)^2-27 m^2 r_g^2-2 \xi \right]}{3 (\delta -2)^5} \,,
\end{equation}
with $\xi=27C_3r_g^2 + (\delta - 2)^2(\delta + 1)$.
So, the maximum of the potential is
\begin{equation}
V(r^{*}_{max})\approx V_{0}L^{2}+V_{1}
\end{equation}
where
\begin{equation}
V_{0}= \frac{\xi}{27 r_g^2} \,, \quad
 V_{1}= \frac{\xi \left( 2\xi+27 m^2 r_g^2\right)}{81 (\delta -2)^2 r_g^2}\,,
\end{equation}
while the higher order derivatives $V^{(k)}(r^{*}_{max})$ for $k=2,...,6$, can be expressed in the following abbreviated manner
\begin{align}
    V^{(2)}(r^{*}_{max})&\approx V_{0}^{(2)}L^{2}+V_{1}^{(2)}\\
    V^{(3)}(r^{*}_{max})&\approx V_{0}^{(3)}L^{2}\\
    V^{(4)}(r^{*}_{max})&\approx V_{0}^{(4)}L^{2}\\
    V^{(5)}(r^{*}_{max})&\approx V_{0}^{(5)}L^{2}\\
    V^{(6)}(r^{*}_{max})&\approx V_{0}^{(6)}L^{2}~.
\end{align}
where
\begin{equation}
\nonumber V_0^{(2)} = -\frac{2 \xi ^2}{729 r_g^4} \,,\quad
\nonumber V_0^{(3)} = \frac{2 \xi^3}{6561 r_g^5} \,, \quad
\nonumber V_0^{(4)} = \frac{16 \xi^3}{19683 r_g^6} \,, \quad 
\end{equation}
\begin{equation}
\nonumber V_0^{(5)} = -\frac{20 \xi^4}{59049 r_g^7} \,, \quad
\nonumber V_0^{(6)} = \frac{4 \xi ^4 (5 \xi -68)}{531441 r_g^8}\,,  \\
\end{equation}
\begin{eqnarray}
\nonumber V_1^{(2)} &=& \frac{4 \xi^3  \left[-(\delta -5) \xi -9 (\delta -2)^2\right]}{6561 (\delta -2)^4 r_g^4} \\
\notag && - \frac{54 \xi^2 m^2 \left[(\delta -5) \xi +3 (\delta -2)^2\right] r_g^2}{6561 (\delta -2)^4 r_g^4} \,. \\
\end{eqnarray}
Moreover, our interest is to evaluate the QNFs for large values of $L$, so we expand the frequencies as power series in $L$. 
It is important to keep in mind that in the eikonal limit, the leading term is linear in $L$.
Next, we consider the following expression in powers of $L$
\begin{equation}
\label{omegawkb}
\omega=\omega_{1m}L+\omega_{0}+\omega_{1}L^{-1}+\omega_{2}L^{-2} + \mathcal{O}(L^{-3})\,,
\end{equation}
where

\begin{equation}
\omega_{1m}=\frac{\sqrt{\xi}}{3 \sqrt{3} r_g}\,,\quad
\omega_{0}= - \frac{i (2 n+1) \sqrt{\xi}}{6 \sqrt{3} r_g}\,, 
\end{equation}

\begin{eqnarray}
\notag \omega_{1} &=& \frac{\sqrt{\xi}}{2592 \sqrt{3} (\delta -2)^2 r_g}\Big[ 3888 m^2 r_g^2-(\delta -4) \delta  \Big((30 n (n+1) \\
&& + 11) \xi +108\Big)-4 (30 n (n+1)-61) \xi -432\Big]\,, 
\end{eqnarray}

\begin{eqnarray}
\notag \omega_2 &=& -\frac{i (2 n+1) \xi ^{3/2}}{{373248 \sqrt{3} (\delta -2)^4 r_g}} \Big[ -72 (5 \delta -34) (5 \delta +14) (\delta -2)^2 \\
\notag && +93312 (\delta -5) m^2 r_g^2+\xi  \Big(\delta  (155 \delta  ((\delta -8) \delta +24)+1952)  \\
&& +235 (\delta -2)^4 n^2+235 (\delta -2)^4 n-32080\Big)\Big]\,. \label{omega2}
\end{eqnarray}\\

Now, considering  the effective potential around the maximum (\ref{expansion}) of the potential and defining the width of the effective potential ($\Delta r^*$) as the interval over which the potential has decayed by a factor of $\epsilon V $ from its maximum value. We obtain:
\begin{equation}
(\Delta r^*)^2 = 2 (1- \epsilon) \frac{V(r^*_{\text{max}})}{- V''(r^*_{\text{max}})} \,.
\end{equation}
Therefore,
\begin{eqnarray} \label{width}
\notag \Delta r^* & \approx & 3 \sqrt{3} r_g \sqrt{\frac{1-\epsilon }{\xi }} \Bigg[1-  \\
\notag && \frac{\xi  \left(2 (\delta -5) \xi +12 (\delta -2)^2+27 (\delta -5) m^2 r_g^2\right)}{18 (\delta -2)^4 L^2}\Bigg] \,. \\
\end{eqnarray}

\begin{figure}[h]
	\begin{center}
		\includegraphics[width=80mm]{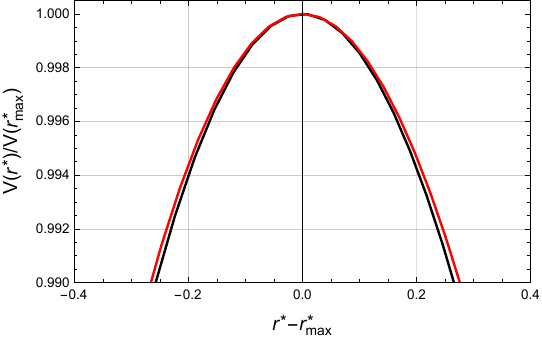}
        \includegraphics[width=80mm]{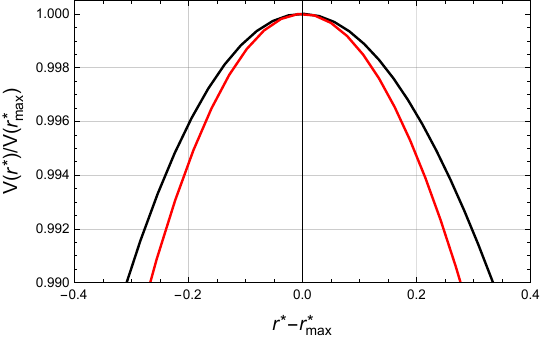}
	\end{center}
	\caption{Effective potential $V(r^*)/V(r^*_{max})$ as a function of $r^*-r^*_{max}$ with $r_g=1$, $\delta = 0.1$ and $C_3 = -0.01$. Top panel for $m=0.1$ and bottom panel for $m=0.5$. Black curves for $\ell=1$ and red curves for $\ell=10$.}
	\label{potencial}
\end{figure}

In Fig. (\ref{potencial}) we show the behavior of the width of the effective potential. Note that for $m=0.1$ the width increases with $\ell$, while for $m=0.5$ the width decreases with $\ell$, which shows an inversion in the behavior of the width, that causes an inverted behavior in $\omega_I$, as we will discuss in the next subsection.

\subsection{Anomalous decay rate behavior}

In this subsection, we will study the behavior of the decaying modes. We expect that the more energetic modes with high angular number $\ell$ would have faster decaying rates. However, the anomalous behavior occurs when the longest-lived modes are those with higher angular numbers, and this can occur with massless and massive probe scalar fields. 
There is a critical mass of the scalar field where the behavior of the decay rate of the QNMs is inverted and can be obtained from the condition $Im(\omega)_{\ell}=Im(\omega)_{\ell+1}$ in the {\it eikonal} limit, that is when $\ell \rightarrow \infty$. The anomalous behavior in the QNFs is possible in asymptotically flat, in asymptotically dS and in asymptotically AdS spacetimes; however, we observed that the critical mass exists for asymptotically flat and for asymptotically dS spacetimes, and it is not present in asymptotically AdS spacetimes for large and intermediate black holes \cite{Aragon:2020tvq}.

The critical mass of the scalar field is given by
\begin{eqnarray}
\label{mcritical}
m_{c}=\frac{\sqrt{\xi \eta  -72 (\delta -2)^2 (5 \delta -34) (5 \delta +14)}}{216 r_g \sqrt{2(5-\delta)}}  \,,
\end{eqnarray}
where 
\begin{equation}
\eta = \delta  \left[155 \delta \left[(\delta -8) \delta +24 \right]+1952 \right]+235 (\delta -2)^4 n(n+1)-32080\,.
\end{equation}

For the fundamental mode $n=0$ and small values of the parameter $\delta$ the critical mass can be approximated to
\begin{eqnarray}
\notag m_c     &&   \approx \frac{\sqrt{5480-541350 C_3 r_g^2}}{540 r_g} -\frac{\delta  \left(4185 C_3 r_g^2+3428\right)}{60 r_g \sqrt{5480-541350 C_3 r_g^2}} \\
\notag && -\frac{\delta ^2 \left(405 C_3 r_g^2 \left(22424625 C_3 r_g^2+26661764\right)+214935296\right)}{3600 \sqrt{10} r_g \left(548-54135 C_3 r_g^2\right){}^{3/2}} \\
\notag && + \mathcal{O}(\delta^3) \,.
\end{eqnarray}
For $\delta=0$, and identifying $C_3$ with an effective cosmological constant $C_3 = - \Lambda_{eff} / 3$ we recover the critical scalar field mass for the Schwarzschild-dS black hole \cite{Aragon:2020tvq}.

\begin{figure}[H]
	\begin{center}
		\includegraphics[width=80mm]{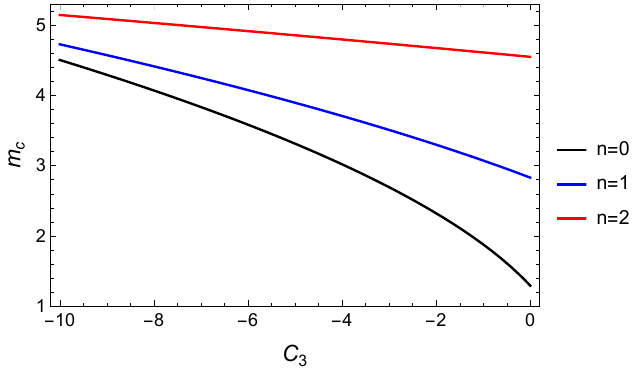}\\
        \includegraphics[width=80mm]{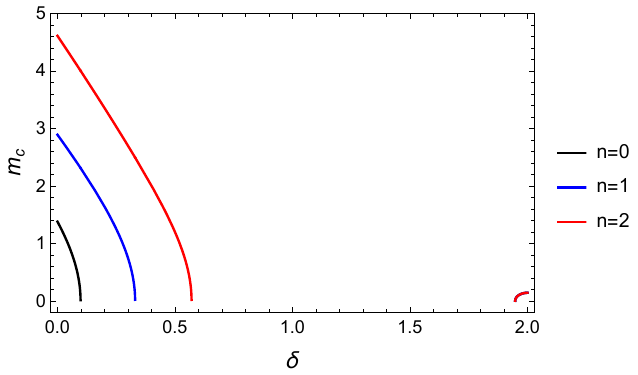}
	\end{center}
	\caption{The behavior of the critical scalar field mass as a function of $C_3$ (top panel), and $\delta$ (bottom panel). Here we have set $r_g=0.1$. Top panel with $\delta=0.01$ and bottom panel with $C_3=-0.01$.}
	\label{f4}
\end{figure}

In Fig. \ref{f4} we plot the behavior of the critical scalar field mass, we can observe that the critical mass decreases when the absolute value of $C_3$ decreases, and the critical mass increases when the overtone number increases (top panel). However, we can observe that there is a range of values of $\delta$, where there is not a critical scalar field mass, so in this range, the longest-lived modes are always those with higher angular number (bottom panel).

\begin{figure}[h]
\begin{center}
\includegraphics[width=0.5\textwidth]{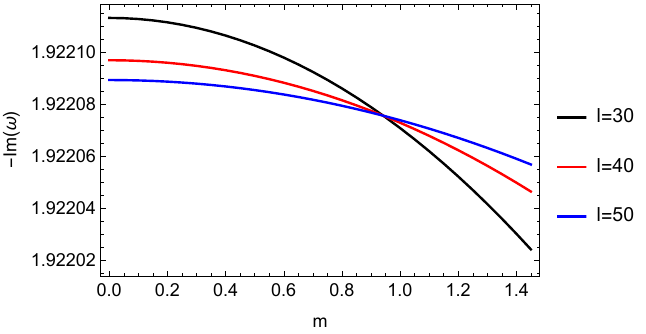}
\end{center}
\caption{The behavior of $-Im(\omega)$ for the fundamental mode ($n=0$) as a function of the scalar field mass $m$ for different values of the angular number $\ell=30,40,50$, with $r_g=0.1$, $\delta=0.05$ and $C_3=-0.01$ using the 6th order WKB method. Here, the WKB method gives $m_{c}\approx0.942$ via Eq. (\ref{mcritical}).}
\label{AB1}
\end{figure}

Now, to illustrate the anomalous behavior, we plot in Fig. \ref{AB1} the behavior of $-Im(\tilde{\omega})$ as a function of $m$, using the 6th-order WKB method. We observe an anomalous decay rate for: $m<m_c$, the longest-lived modes correspond to the highest angular number $\ell$, while for $m>m_c$, the longest-lived modes correspond to the lowest angular number.

\section{Connecting the Lyapunov exponent with the  anomalous behavior}
\label{CLmc}

For a general static and spherically symmetric metric:

\begin{equation}
ds^2 = f(r) dt^2 - \frac{1}{f(r)}dr^2-a^2(r) d \Omega^2 \,,
\end{equation}
the radial geodesic equation is given by Eq. (\ref{motion}), with an effective potential
\begin{equation}
V_{\text{eff}} = f(r) \left( m^2 + \frac{L^2}{a^2(r)}{} \right) \,.
\end{equation}

Expanding the potential around its maximum --which corresponds to the unstable circular orbits (the photon sphere for massless particles)-- we get
\begin{equation} \label{taylor}
V_{\text{eff}} (r) = V_{\text{eff}} (r_{\text{max}}) + \frac{1}{2!} V''_{\text{eff}}(r_{\text{max}}) (r- r_{\text{max}})^2 + \dots
\end{equation}
We define the width of the effective potential for geodesics, $\Delta r_G$, as the interval over which the potential falls by a fraction $\epsilon V_{\text{eff}}$ from its maximum value. From the Taylor expansion above, it follows that
\begin{equation} \label{deltar}
(\Delta r_G)^2 = 2 (1- \epsilon) \frac{V_{\text{eff}}(r_{\text{max}})}{- V_{\text{eff}}''(r_{\text{max}})} \,.
\end{equation}

Additionally, the condition $V_r = 0$ for circular geodesics (applicable to both massless and massive particles) implies $V_{\text{eff}}(r_{\text{max}}) = E^2$. By introducing the tortoise coordinate, defined by $d r^* = dr / f(r)$, the width in this coordinate becomes $\Delta r^*_G = \Delta r_G / f(r_{\text{max}})$, resulting in

\begin{equation}
\Delta r^*_G = \frac{\sqrt{2 (1 - \epsilon)}E}{ f(r_{\text{max}}) \sqrt{-V''_{\text{eff}}(r_{\text{max}})}} \,.
\end{equation}

Besides, the Lyapunov exponent, given in Eq. (\ref{sle}), valid in the same general static, spherically symmetric spacetime, can be expressed as

\begin{equation}
\lambda = \pm \sqrt{\frac{-V''_{\text{eff}}(r_{\text{max}})}{2}} \frac{f(r_{\text{max}})}{E} \,.
\end{equation}

Hence, we arrive at the following relation
\begin{equation} \label{relation}
 \Delta r^*_G = \frac{\sqrt{1- \epsilon}}{\lambda} \,.
\end{equation}
This result demonstrates that the Lyapunov exponent is inversely proportional to the width of the effective potential for unstable circular geodesics in general static, spherically symmetric spacetimes. This relationship underscores the sensitivity of chaotic dynamics to the spatial extent of the potential well: although these orbits are by nature unstable, a broader potential well corresponds to weaker instability (i.e., a longer divergence timescale) than a narrowly confined potential.

As previously studied, an anomalous behavior in the decay rates of QNMs occurs when the longest-lived modes correspond to higher angular numbers. This phenomenon has been observed for massless and massive probe scalar fields. A critical mass exists for the scalar field, beyond which the decay rate behavior inverts. This inversion can be identified by the condition $Im(\omega)_{\ell}=Im(\omega)_{\ell+1}$ in the {\it eikonal} limit; that is, when $\ell \rightarrow \infty$. This inverted behavior in the imaginary part of the frequencies is due to inversion of the behavior of the width of the effective potential of the scalar field.

Therefore, it is pertinent to investigate whether, in the limit where the potentials governing the geodesics and the scalar field converge, the anomalous behavior of QNMs is associated with the Lyapunov exponent. This examination could provide deeper insights into the interplay between geodesic stability and the decay rates of QNMs, particularly at the limit where the angular number $\ell$ becomes large.

\subsubsection{Massless particles}

Note that in the eikonal limit the effective potential for the geodesics is the same as that for the scalar field for massless particles. So, considering Eq. (\ref{deltar}), with
\begin{equation}
 V_{\text{eff}}(r_{ps}) =  \frac{L^2\xi}{27 r_g^2}\,,\quad 
 V_{\text{eff}}''(r_{ps}) = - \frac{2 L^2 (\delta-2)^4}{81 r_g^4}\,,
\end{equation}
we obtain
\begin{equation}
\Delta r_G =  \frac{  3r_g (1- \epsilon) \xi^{1/2}}{(\delta-2)^2}\,. 
\end{equation}
Now, using the tortoise coordinate, the above expression can be written as $\Delta r^*_G = \Delta r_G / f(r_{ps})$, where $f(r_{ps}) = \frac{\xi}{3 (\delta-2)^2}$, which yields
\begin{equation}
\Delta r^*_G = \frac{3 r_g\sqrt{3 (1- \epsilon )}  }{\sqrt{\xi}}\,.
\end{equation}
Therefore, considering (\ref{LE}) the above expression can be written as
\begin{equation} \label{rrgg}
 \Delta r^*_G = \frac{\sqrt{1- \epsilon}}{\lambda_0}\,.
\end{equation}
So, the Lyapunov exponent is inversely proportional to the width of the effective potential for null geodesics, in concordance with the general result (\ref{relation}).

Considering Eq. (\ref{sle}) and Eq. (\ref{LE}), the QNFs (\ref{omegawkb}) in the eikonal limit can be written as

\begin{equation}
    \omega= \lambda_0 L - i\left(n+\frac{1}{2}\right)\abs{\lambda_0}\,,
\end{equation}
or
\begin{equation}
    \omega= \Omega_{0c} L - i\left(n+\frac{1}{2}\right)\abs{\lambda_0}\,,
\end{equation}
where $\Omega_{0c}=\dot{\varphi}/\dot{t}=(f^\prime(r_{ps})/2r_{ps})^{1/2}$ is the orbital angular velocity for null geodesics. This relation connects the quasinomal frequencies $\omega$ with the Lyapunov exponent $\lambda_0$.  
This is similar to the relation (46) obtained in \cite{Cardoso:2008bp} in which the QNMs of any spherically symmetric, asymptotically flat spacetime are given by the frequency and instability timescale of the unstable circular null geodesics expressed by the Lyapunov exponent. Note that in our case, the spacetime is asymptotically dS-like spacetime for $C_3 < 0$.

\subsubsection{Massive particles}

For massive particles in the eikonal limit, the effective potential governing geodesics coincides with that of a scalar field, provided the condition $\frac{f'(r)}{m^2 r}<<1$ holds outside the horizon. Evaluating this at the horizon radius gives $\frac{f'(r_H)}{m^2 r_H}<<1$, which leads to the condition
\begin{equation}
\frac{(3 r_g+(\delta -2) r_H) (3 r_g+\delta  r_H)}{3 m^2 r_g r_H^3} << 1 \,.
\end{equation}

For massive particles, the effective potential $V_{\text{eff}}$ is defined by (\Ref{n2}).
In this case, the width of the potential can be expanded using the tortoise coordinate, as

\begin{equation} \label{dr}
\Delta r^*_G = \frac{3 r_g\sqrt{3 (1-\epsilon)}}{\sqrt{\xi}} \left[1-\frac{3 (\delta -5) m^2 r_g^2 \xi}{2 (\delta -2)^4 L^2} + \dots \right]  
\end{equation}

\begin{figure}[H]
	\begin{center}
		\includegraphics[width=80mm]{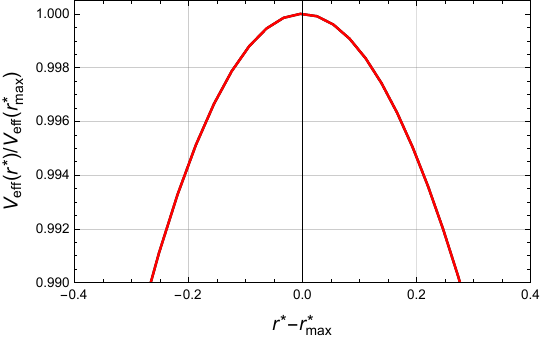}
        \includegraphics[width=80mm]{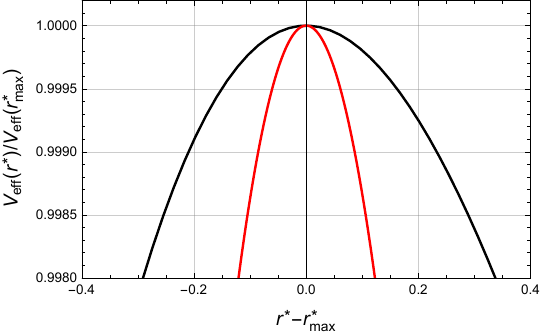}
	\end{center}
	\caption{Effective potential $V_{\text{eff}}(r^*)/V_{\text{eff}}(r^*_{\text{max}})$ as a function of $r^*-r^*_{\text{max}}$ with $r_g=1$, $\delta = 0.1$ and $C_3 = -0.01$. Top panel for massless particles and bottom panel for massive particles. Black curves for $L=2$ and red curves for $L=20$. In the top figure the width does not change with $L$.}
	\label{potencialg}
\end{figure}

This width approaches the width of the effective potential of the Schr\"{o}dinger equation for the scalar field (\ref{width}) when $m$ becomes dominant in the second term of that formula $\Delta r^*_G \rightarrow \Delta r^*$. In Fig. (\ref{potencialg}) we show the behavior of the width of the effective potential of geodesics. We observe a different behavior for massless and massive particles. For massless particles, the width does not change with $L$, while for massive particles the width decreases with $L$, this last behavior is similar to that of the width of the potential of the scalar field.


On the other hand, from (\ref{LEm}) we obtain
\begin{equation} \label{lm}
 \lambda_m =  \frac{\sqrt{\xi}}{3 \sqrt{3} r_g} \left[ 1 +  \frac{3 (\delta -5) m^2 r_g^2 \xi}{2 (\delta -2)^4 L^2}+\dots \right]
\end{equation}

Therefore, comparing (\ref{dr}) and (\ref{lm}) we find

\begin{equation}
\Delta r^*_G = \frac{\sqrt{1-\epsilon}}{\lambda_m} \,.
\end{equation}

If the mass term dominates in the imaginary part $\omega_2$ of the QNFs, we can write up to the third order beyond the eikonal limit

\begin{equation}
\notag Im(\omega) = -\frac{\sqrt{\xi}}{6 \sqrt{3} r_g}\Bigg[1+ \\
 \frac{3 (\delta -5) m^2 r_g^2 \xi}{2 (\delta -2)^4 L^2}+\dots \Bigg]\,.
\end{equation}

Now, comparing this expression with that of the Lyapunov exponent (\ref{lm}), we find

\begin{equation}
Im (\omega) = - \frac{1}{2} \lambda_m\,.
\end{equation}
Therefore, in the regime where the scalar field mass dominates in Eq. (\ref{omega2}), the imaginary part of the QNFs is proportional to the Lyapunov exponent beyond the eikonal limit.

Moreover, the angular velocity is given by
\begin{equation}
\Omega = \frac{\dot{\varphi}}{\dot{t}} = \left( \frac{f'}{2r}  \right)^{1/2}\,.
\end{equation}
So, the angular velocity of the unstable circular timelike orbit is given by 
\begin{equation} \label{angular}
\Omega_{c} \approx \frac{\sqrt{\xi}}{3 \sqrt{3} r_g} \left[1-\frac{9 m^2 r_g^2}{2 (\delta -2)^2 L^2}\right]\,.
\end{equation}
In addition, the real part of the QNFs up to third order beyond the eikonal limit, when the mass term dominates in $\omega_1$, is given by
\begin{equation} \label{real}
Re(\omega) \approx \frac{\sqrt{\xi} L }{3 \sqrt{3} r_g}  \left[1+\frac{9 m^2 r_g^2}{2 (\delta -2)^2 L}\right]\,.
\end{equation}

Consequently, the relation $Re(\omega) = \Omega_c \ell$ holds in the eikonal (high-$\ell$) regime for massive particles, but it breaks down beyond this limit due to the opposite sign of the sub-leading corrections in expressions (\ref{angular}) and (\ref{real}). This shows that the angular velocity of the unstable circular orbit decreases with the particle mass $m$, while the real part of the QNFs increases with $m$. \\

Furthermore, the critical scalar field mass (\ref{mcritical}), defined in the eikonal limit, can be written as a function of the Lyapunov exponent $\lambda_0$. Note that in the eikonal limit $\lambda_m(L \rightarrow \infty)  = \lambda_0$ and
\begin{eqnarray}
\notag m_c &=& \frac{1}{24 \sqrt{6(5 - \delta) (\delta +1)}}\Big[72 C_3 (5 \delta -34) (5 \delta +14)+ \\
\notag &&( 155 \delta ^5-1085 \delta ^4+ 
  2480 \delta ^3+3872 \delta ^2-22928 \delta +\\
&&  2192) \lambda _0^2 \Big]^{1/2}\,.
\end{eqnarray}
It is worth mentioning that for $\delta = \delta_c \approx 0.0972$, the critical mass does not vary with $\lambda_0$. For $ \delta < \delta_c$ the  critical scalar field mass increases with $\lambda_0$, while for $\delta > \delta_c$ the critical scalar field mass decreases with $\lambda_0$. When $C_3=0$ the critical mass is proportional to $\lambda_0$ (and according to (\ref{rrgg}) is inversely proportional to $\Delta r^*_G$) and exists for $\delta < \delta_c$. For $\delta=0$ reduces to
\begin{equation}
m_c = \frac{\sqrt{2192 \lambda _0^2-34272 C_3}}{24 \sqrt{30}} \,,
\end{equation}
and for Schwarzschild ($C_3=0$) we find $m_c \sim \lambda_0$.

\begin{figure}[h]
\begin{center}
\includegraphics[width=0.4\textwidth]{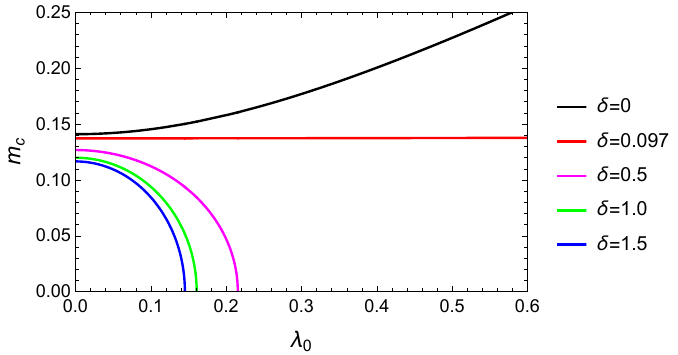}
\end{center}
\caption{The behavior of the critical scalar field mass $m_c$ for the fundamental mode ($n=0$) as a function of the Lyapunov exponent $\lambda_0$ for different values of $\delta=0, 0.0972,0.5,1.0,1.5$ and $C_3=-0.01$.}
\label{AB11}
\end{figure}

\begin{figure}[h]
\begin{center}
\includegraphics[width=0.4\textwidth]{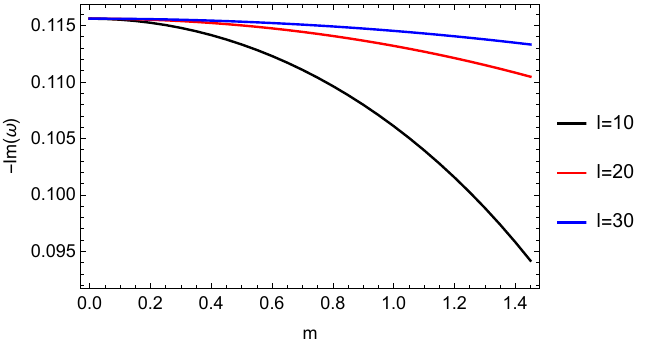}
\end{center}
\caption{The behavior of $-Im(\omega)$ for the fundamental mode ($n=0$) as a function of the scalar field mass $m$ for different values of the angular number $\ell=10,20,30$, with $r_g=1$, $\delta=1$ and $C_3=-0.0206302$ using the 6th order WKB method. Here, the WKB method gives $m_{c}\approx0$.}
\label{AW}
\end{figure}

In Fig. \ref{AB11}, we show the behavior of the critical scalar field mass as a function of the Lyapunov exponent. We observe that the critical mass decreases when the Lyapunov exponent increases and when the parameter $\delta$ increases for a fixed value of the Lyapunov exponent. Note that for $ \delta > \delta_c$ there is a value of the Lyapunov exponent for which the critical scalar field mass is null, given by
{\small
\begin{eqnarray}
\notag \tilde{\lambda}_0 &=&  \frac{\pm 6 \sqrt{2}\Big[-C_3 (5 \delta -34) (5 \delta +14) \Big]^{1/2}}{\sqrt{155 \delta ^5-1085 \delta ^4+ 2480 \delta ^3+3872 \delta ^2-22928 \delta +2192}}\,,\\
\end{eqnarray}}
for $\lambda_0 = \tilde{\lambda}_0$ the anomalous behavior of the decay rate is avoided, and the longest-lived modes are the ones with the smallest angular number, which is shown in Fig. \ref{AW} for $\lambda_0=\pm 4\sqrt{29/8682}=\pm 0.231179$, for $\delta=1$, $r_g=1$, $C_3=-806/39069=-0.0206302$. However, for $\delta < \delta_c$, the anomalous behavior cannot be avoided. 

\newpage

\section{Conclusions}
\label{conc}

We studied the stability of the Weyl geometry considering a specific black hole solution discussed in \cite{Yang:2022icz,Sakti:2024pze}. In the Weyl geometric black hole solutions extra terms appear due to the presence of a vector field and a scalar field. The presence of the scalar field can be understood as describing the energy density and pressure of an effective fluid dressing the Weyl geometric black hole by a material cloud. Therefore, in the Weyl geometry, the black hole solutions contain scalar hair, which determines the strength of the gravitational interaction.

We calculated the geodesics of massless and massive test scalar fields orbiting outside the background Weyl black hole. Using the Lyapunov exponent  we constrained the values of the vector field and the scalar field, which appeared in the Weyl black hole. Motivated by the fact that QNMs can be interpreted as particles trapped in unstable circular geodesics and slowly leaking out \cite{Cardoso:2008bp} we calculated the QNMs of a test scalar field perturbing the Weyl black hole. Then, as it is known that the leak timescale is
given by the principal Lyapunov exponent, we find a relation connecting the quasinormal frequencies $\omega$ with the Lyapunov exponent $\lambda_0$ and this relation is valid in an asymptotically dS-like spacetime and asymptotically flat spacetime. We also studied the connection of the Lyapunov exponent to the anomalous decay of QNMs.

In conclusion, Weyl black holes provide a compelling framework for investigating deviations from classical GR. Their unique properties, particularly the presence of vector and scalar fields, emphasize the richness of modified gravity theories and their potential to explain cosmological phenomena that remain elusive in standard paradigms. These findings not only deepen our understanding of black hole physics and their stability but also open new avenues for probing the fundamental nature of spacetime and gravity. Besides improving our physical understanding of ringdown radiation, a deeper exploration of this analogy could have important implications to the interpretation of black hole binary mergers and their use in gravitational-wave data analysis.

Moreover, the findings highlight the necessity of revisiting key astrophysical phenomena in the context of Weyl black holes. For instance, accretion dynamics, quasi-periodic oscillations, and other observational signatures should be explored under the modified spacetime structure. A possible chaotic behavior near the horizons of these black holes opens a window into understanding the interplay between geometry and dynamics in non-classical spacetimes.

Future research should aim to bridge the gap between theoretical predictions and observational prospects. This includes refining the models to account for the effects of Weyl fields on high-energy astrophysical processes and identifying potential observational signatures that could confirm the presence of Weyl black holes. Additionally, exploring the thermodynamic properties and radiation mechanisms in these spacetimes could shed light on their role as probes of alternative gravity theories.

\section{Acknowledgements}

The authors would like to thank the anonymous referee for valuable comments and suggestions that helped to improve the clarity and quality of this manuscript. We also wish to thank Kyriakos Destounis, Ioannis Gialamas and  Dimitris Giataganas for valuable discussions and suggestions.



\end{document}